\documentclass{article}
\usepackage{amsmath,epsfig}
\usepackage{multirow}
\usepackage{booktabs}
\usepackage{url}
\usepackage[preprint]{spconfa4}

\let\OLDthebibliography\thebibliography
\renewcommand\thebibliography[1]{
  \OLDthebibliography{#1}
  \setlength{\parskip}{0pt}
  \setlength{\itemsep}{0pt plus 0.3ex}
}

\begin{document}\sloppy

% Example definitions.
% --------------------
\def\x{{\mathbf x}}
\def\L{{\cal L}}

% Title.
% ------
\title{Deep Learning based Full-reference and No-reference Quality Assessment Models for Compressed UGC Videos}
%
% Address.
% ---------------
\name{Wei Sun, Tao Wang, Xiongkuo Min, Fuwang Yi, and Guangtao Zhai \thanks{This work was supported by the National Natural Science Foundation of (No.61901260, 61831005, 61831015, 61771305, and U1908210).}}
\address{Institue of Image Communication and Information Processing, Shanghai Jiao Tong University, China \\
Email: sunguwei@sjtu.edu.cn}

\maketitle

\begin{abstract}
In this paper, we propose a deep learning based video quality assessment (VQA) framework to evaluate the quality of the compressed user's generated content (UGC) videos. The proposed VQA framework consists of three modules, the feature extraction module, the quality regression module, and the quality pooling module. For the feature extraction module, we fuse the features from intermediate layers of the convolutional neural network (CNN) network into final quality-aware feature representation, which enables the model to make full use of visual information from low-level to high-level. Specifically, the structure and texture similarities of feature maps extracted from all intermediate layers are calculated as the feature representation for the full reference (FR) VQA model, and the global mean and standard deviation of the final feature maps fused by intermediate feature maps are calculated as the feature representation for the no reference (NR) VQA model. For the quality regression module, we use the fully connected (FC) layer to regress the quality-aware features into frame-level scores. Finally, a subjectively-inspired temporal pooling strategy is adopted to pool frame-level scores into the video-level score. The proposed model achieves the best performance among the state-of-the-art FR and NR VQA models on the Compressed UGC VQA database and also achieves pretty good performance on the in-the-wild UGC VQA databases.
\end{abstract}
\section{Introduction}
\label{sec:intro}

With the rapid development of mobile devices and wireless networks in recent years, watching and sharing user's generated content (UGC) videos on social media applications has been a popular daily activity for contemporary people. To reduce the cost of video storage and transmission, the service providers will first compress UGC videos before they are transmitted to the user client, which inevitably degrades the quality of UGC videos. Therefore, it is necessary to develop an effective video quality assessment (VQA) model to evaluate the quality of compressed UGC videos, which on one hand can help the service provider recommend high-quality videos to users, and on the other hand can guide the development of more effective compression algorithms.

Although objective VQA algorithms have been studied for many years, most of them are developed for professionally generated content (PGC) videos, which are shot by photographers using the professional photographic equipment. As a result, the pristine PGC videos are usually of high quality and the quality of compressed PGC videos mainly depends on the degree of distortions introduced during compression and transmission. However, UGC videos are usually captured by amateur users using smartphone cameras under various shooting environments. Therefore, the quality of pristine UGC videos is affected by the video content such as its aesthetic characteristics and authentic distortions such as low visibility, noise, jitter, etc. The compressed UGC videos are further degraded by compression artifacts, which poses a challenge to existing VQA models.

\begin{figure*}[!t]
	\centering
	\includegraphics[height=2.4in]{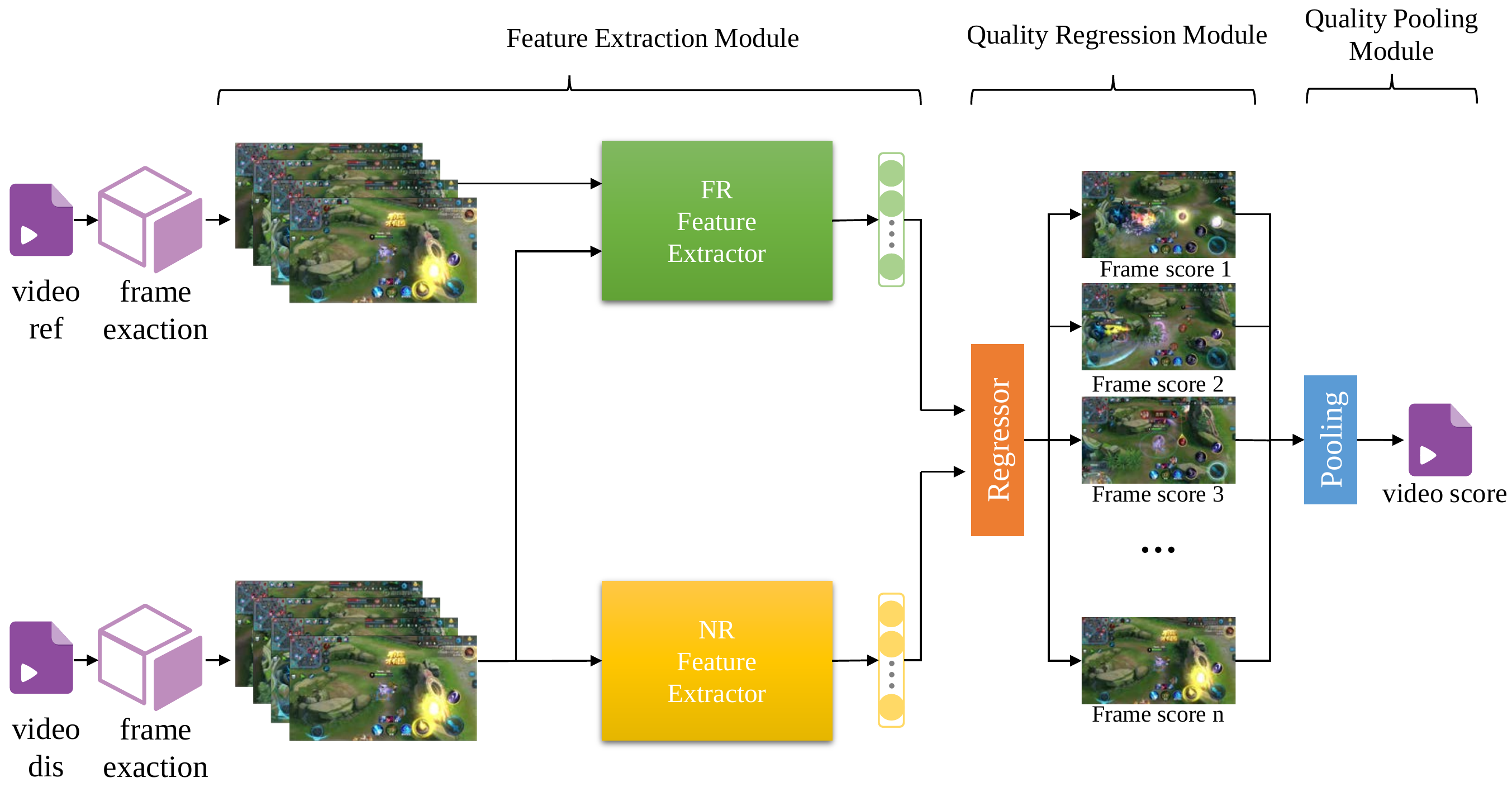}
	\caption{The proposed FR and NR VQA framework. The framework includes the feature extraction module, the quality regression module, and the quality pooling module.}
	\label{framework}
	\vspace{-0.3cm}
\end{figure*}

Generally speaking, objective VQA can be divided into full-reference VQA (FR VQA), reduced-referenced IQA (RR VQA), and no-reference (NR VQA) according to whether to access the reference information. FR VQA and RR VQA models require full and part reference video information respectively, while NR VQA models only take the distorted video as the input, which is more difficult but is also more practical in real applications.

%Due to the full access of the reference information.
%Vu el al. consider human perception on motion distortions and then incorporate it to the most apparent distortion (MAD) index to obtain spatial-temporal MAD (ST-MAD) \cite{vu2011spatiotemporal}.
%Seshadrinathan and Bovik develop the Motion-based Video Integrity Evaluation (MOVIE) index \cite{seshadrinathan2009motion} by evaluating motion quality along computed motion trajectories.
FR VQA models usually measure the fidelity between the reference and distorted frames of videos as the video quality. The most widely used fidelity metrics are PSNR and SSIM \cite{wang2004image}, which respectively calculate the pixel-wised Euclidean distance and structural similarity between each reference and distorted frame. Then, the average pooling strategy is usually adopted to pool them into the video score.  However, the perceptual video quality is also related to the temporal information such as the motion, and these image quality assessment (IQA) \cite{zhai2020perceptual} based methods cannot model the quality-aware feature of the time domain. Therefore, some studies try to extend existing IQA methods to the video domain by incorporating the temporal information. For example, Moorthy and Bovik propose a motion compensated SSIM (MC-SSIM) \cite{moorthy2010efficient} by evaluating structural retention between motion-compensated regions. Vu \textit{et al.} consider human perception on motion distortions and then incorporate it to the most apparent distortion (MAD) index to obtain spatial-temporal MAD (ST-MAD) \cite{vu2011spatiotemporal}. Recently, learning based methods show great abilities in the VQA field. Video Multi-method Assessment Fusion (VMAF) extracts multiple IQA features as well as motion features and learns a Support Vector Regressor (SVR)  to map these features into the video quality. DeepVQA \cite{kim2018deep} evaluates the spatio-temporal visual quality via a convolutional neural network (CNN) and a convolutional neural aggregation network (CNAN). C3DVQA \cite{xu2020c3dvqa} utilizes the 3D CNN to learn spatio-temporal features for video quality evaluation.

%NR VQA is more difficult but is also more practical in real applications since it does not need any additional information.
%, which is the same with the mentioned FR VQA methods 
A naive NR VQA method is to compute the quality of each frame via popular NR IQA methods such as NIQE \cite{mittal2012making}, BRISQUE \cite{mittal2012no}, etc., and then pool them into the video quality score. A comparative study of various temporal pooling strategies on popular NR IQA methods can refer to \cite{tu2020comparative}. As discussed above, the temporal information is also important for VQA. V-BLIINDS \cite{saad2014blind} is a spatio-temporal natural scene statistics (NSS) model for videos by quantifying the NSS feature of frame-differences and motion coherency characteristic. TLVQM \cite{korhonen2019two} extracts abundant spatio-temporal features such as motion, jerkiness, blurriness, noise, blockiness, color, etc. at two levels of high and low complexity. VIDEVAL \cite{tu2021ugc} further combines the selected features from typical NR I/VQA methods to train a SVR model to regress them into the video quality. Since video content also affects its quality, especially for UGC videos, understanding the video content is beneficial to NR VQA. Hence, some studies \cite{li2019quality,chen2020rirnet} extract semantic-level features of each frame using the pretrained CNN model and then use the sequential modeling method such as recurrent neural network (RNN) to learn the temporal relationship of each frame and regress them into the video score.

Though lots of VQA models have been proposed in the literature, there are still some critical issues that need to be considered. First, previous studies \cite{gao2017deepsim}\cite{sun2019mc360iqa} have demonstrated that both low-level visual features and high-level semantic information affect the quality of images/videos. However, most existing VQA models only consider one of these factors. For example, traditional hand-crafted feature based models \cite{saad2014blind}\cite{korhonen2019two} extract the low-level visual features, but it is difficult to understand the video content, while recent deep learning based methods  \cite{li2019quality}\cite{chen2020rirnet} utilize the pretrained semantic features but ignore the effect of low-level visual features. Second, most existing VQA models are two stage methods, which first extract hand-craft features or deep semantic features and then regress them into the video quality via machine learning methods like SVR, RNN, etc. It lacks an end-to-end learning method to learn the relationship between the video quality and raw pixels of video frames. 

In this paper, we propose an effective deep learning based VQA framework, which allows it to be trained in an end-to-end manner. The proposed framework is illustrated in Fig. \ref{framework}, which contains three modules, the feature extraction module, the quality regression module, and the quality pooling module. For the feature extraction module, we fuse the features from intermediate layers of the CNN network into final feature representation, which makes the model take full advantage of visual information from low-level to high-level. Specifically, for the FR VQA task, we calculate the structure and texture similarities of each layer of the CNN network as the quality-aware feature representation. For the NR VQA task, we hierarchically add the feature maps from intermediate layers into the final feature maps and calculate their global mean and standard deviation as the quality-aware feature representation. We use the fully connected (FC) layer to regress the quality-aware features into the frame-level quality score. Finally, a subjectively-inspired temporal pooling strategy is adopted to obtain the final video quality. The experimental results on the Compressed UGC VQA database show that the proposed model outperforms other state-of-the-art VQA models by a large margin.

\section{Proposed Method}

The framework of the proposed model is illustrated in Fig. \ref{framework}, which consists of the feature extraction module, the quality regression module, and the quality pooling module. Since the adjacent frames of the video contain lots of redundancy information, we just extract $N$ frames from each video to calculate their quality-aware feature representation via the feature extraction module. Then the quality regression module is used to map the quality-aware features into frame-level quality scores. Finally, we perform a subjectively-inspired temporal pooling strategy to obtain the video quality score.

\begin{figure}[!t]
	\centering
	\includegraphics[height=1.5in]{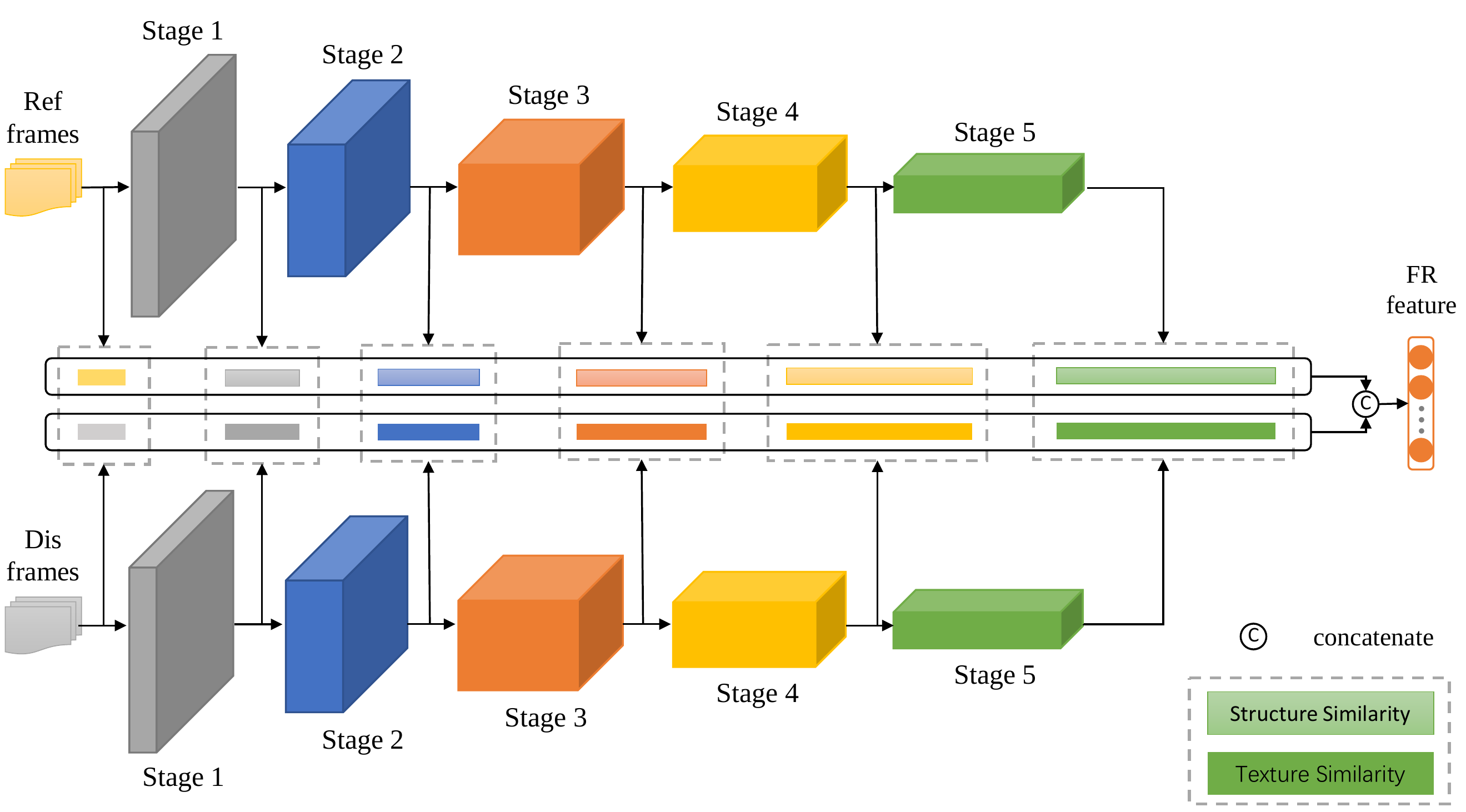}
	\caption{The network architecture of the FR feature extraction module.}
	\label{FR}
	\vspace{-0.3cm}
\end{figure}

\subsection{Feature Extraction Module}
Since the features extracted from different stages of a CNN model represent different visual information \cite{zeiler2014visualizing}, recently proposed IQA models try to fuse the features extracted from intermediate layers of the CNN model, which can make the model fully utilize the visual information from low-level to high-level and learn better quality-aware feature representations. We also follow this routine to design our FR and NR feature extraction module.

\subsubsection{FR Feature Extraction Network}
As stated in Introduction Section, FR VQA models are actually fidelity metrics that can reflect the perceptual quality. Traditional VQA models calculate the structural similarities of two images as the quality, but they are not always consistent with human perception. \cite{zhang2018unreasonable} finds that deep features are very effective to represent the quality characteristics and they propose the perceptual similarity which calculates the L2 distance of feature maps of each pair extracted by intermediate layers of a CNN model. Inspired by the form SSIM, \cite{ding2020image} proposes the deep structure and texture similarities through computing the global means and the global correlation of feature maps of each pair. In this paper, we also adopt this measure to evaluate the quality distance between two feature maps.
%Specifically, we use the ResNet-50 \cite{he2016deep} as the backbone of the feature extraction network. We calculate the structure and texture similarity of feature maps extracted by all stages, i.e. the layer $conv1$, $conv2\_3$, $conv3\_4$, $conv4\_6$, $conv5\_3$ as well as their inputs between reference and distorted frames. The structure and texture similarity are respectively defined as:
Assume that there are $N_s$ stages in a CNN model, and $F_{i}$ is the feature maps extracted from the $i$-th stage, where $i\in [1, 2, ..., N_s]$. $F_{i}^{j}$ is the $j$-th feature map in $F_{i}$. Then, the structure and texture similarities are defined as:
\begin{equation}
\begin{aligned}
f_{texture}\left(F_{r,i}^{j}, F_{d,i}^{j}\right) &=\frac{2 \mu_{F_{r,i}^{j}} \mu_{F_{d,i}^{j}}+c_{1}}{(\mu_{F_{r,i}^{j}})^2+(\mu_{F_{d,i}^{j}})^2+c_{1}}, \\
f_{strcture}\left(F_{r,i}^{j}, F_{d,i}^{j}\right) &=\frac{2 \sigma_{F_{r,i}^{j} F_{d,i}^{j}}+c_{2}}{(\sigma_{F_{r,i}^{j}})^2+(\sigma_{F_{d,i}^{j}})^2+c_{2}},
\end{aligned}
\end{equation}
where $ \mu_{F_{r,i}^{j}}$, $\mu_{F_{d,i}^{j}}$, $(\sigma_{F_{r,i}^{j}})^2$, $(\sigma_{F_{d,i}^{j}})^2$, and $\sigma_{F_{r,i}^{j} F_{d,i}^{j}}$ are the global means and variances of feature maps $F_{r,i}^{j}$ and $F_{d,i}^{j}$, and the global covariance between  $F_{r,i}^{j}$ and $F_{d,i}^{j}$, respectively. $F_{r,i}^{j}$ and $F_{d,i}^{j}$ are the feature maps of the reference and distorted frames respectively. $c_{1}$ and $c_{2}$ are the small constants to avoid numerical instability. Finally, we obtain the structure and texture similarities of feature maps extracted by each stage of the CNN model and all of them constitute the quality-aware feature representation of our FR VQA model $f_{fr}$. We illustrate the FR feature extraction network in Fig. \ref{FR}.

\begin{figure}[!t]
	\centering
	\includegraphics[height=1.5in]{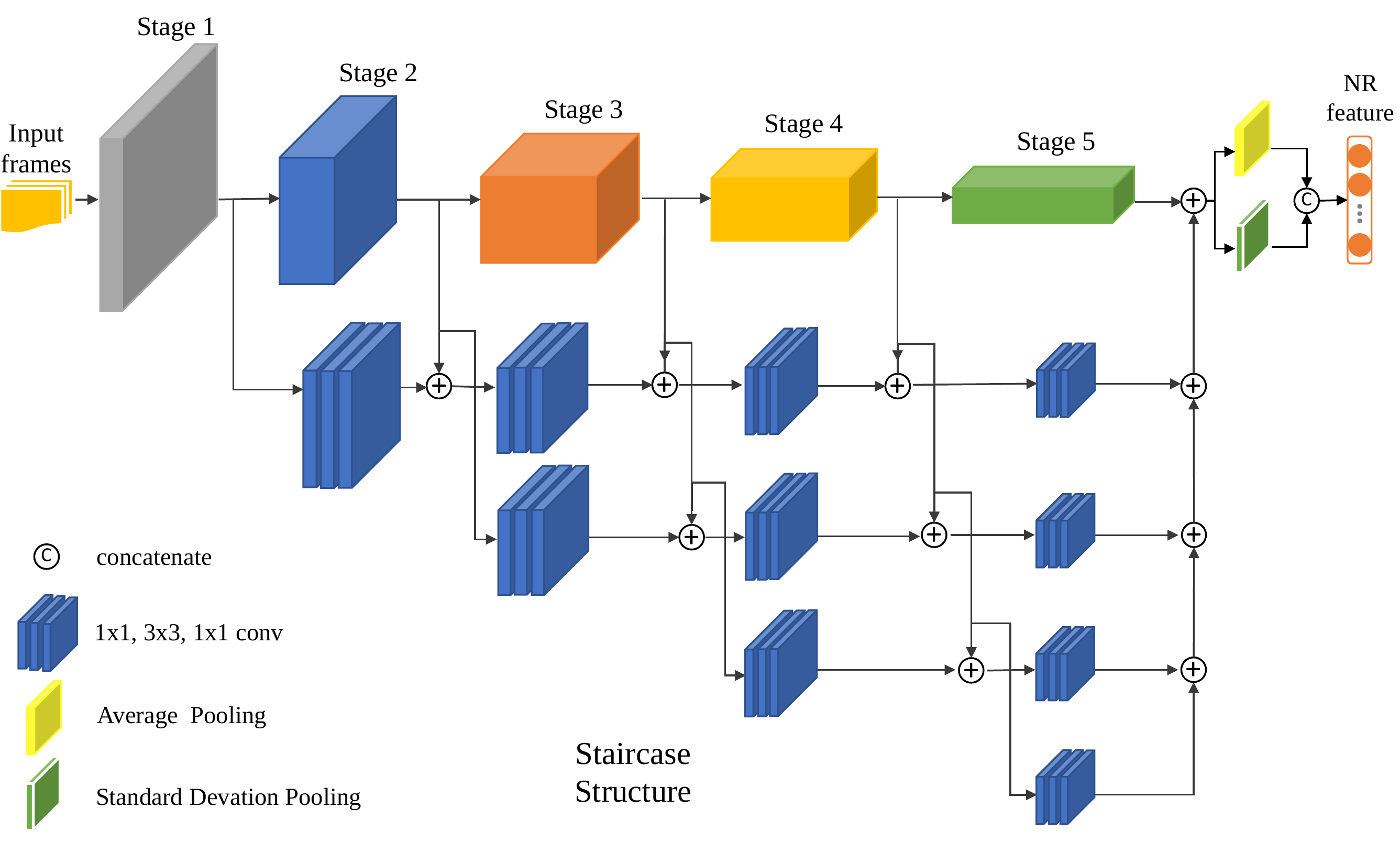}
	\caption{The network architecture of the NR feature extraction module.}
	\label{NR}
	\vspace{-0.3cm}
\end{figure}

\subsubsection{NR Feature Extraction Network}
Since the NR feature extraction module needs to obtain the features that can represent distortions and content just using the distorted frames, we utilize the recently proposed staircase structure \cite{sun2021blind} to obtain the quality-aware feature representation of the NR VQA model, which considers both the image distortion and content. The staircase structure is illustrated in Fig. \ref{NR}, which hierarchically integrates the features from intermediate layers into the final feature representation. To be more specific, the feature maps $F_{1}$ will be continually added to the feature maps $F_{i}$ until to the final feature maps $F_{N_s}$ to derive $\widetilde{F}_{1}$, where $1 < i < N_s$. Then we do the same operation on the feature maps $F_i$ to derive $\widetilde{F}_{i}$. Hence, the final fusing feature maps $F_{s}$ is calculated as:

\begin{eqnarray}
\begin{aligned}
\label{feature_fuse3}
& F_s = \sum^{N_{s}-1}_{i=1} \widetilde F_i + F_{N_{s}} .
\end{aligned}
\end{eqnarray}

For most popular CNN models, the dimension of the feature maps at the current stage is half that of the previous stage while the number of channels is twice that of the previous stage. To make the number of channels and the dimension of feature maps at different stages the same, we introduce a bottleneck structure consisting of three convolution operations to downscale the dimension and increase the channels. Specifically, we first reduce the channels of feature map $F_i$ to a quarter through the 1$\times$1 convolution layer to decrease the computation complexities of the whole procedure. Then we utilize the 3$\times$3 convolution layer with a stride of 2 to reduce the resolution of $F_i$ to half. Finally, $F_i$ is passed through the 1$\times$1 convolution layer to increase the number of channels for eight times.

\begin{table}
	\small
	%\footnotesize
	\centering
	\renewcommand{\arraystretch}{1.15}
	\caption{The performance of the proposed FR VQA model and the compared models on the validation set of the Compressed UGC VQA database.}
	\label{FR_Compressed}
	%\resizebox{1\textwidth}{!}{
	\begin{tabular}{c|cccc}
		\toprule[.15em]
		
		Methods & SROCC & KROCC & PLCC & RMSE   \\
		\hline
		
		SSIM \cite{wang2004image}    &0.8385 &0.6395  & 0.8375  &0.6934  \\
		
		%ILNIQE     &0.903 &0.824  & 0.552 &0.904    \\
		LPIPS \cite{zhang2018unreasonable} &0.8870  	&0.7080 &0.9191 &0.4999\\
		DISTS\cite{ding2020image} &0.9013 &0.7330 &0.9424 &0.4244 \\
		C3DVQA \cite{xu2020c3dvqa} & 0.9054 	&0.7365& 0.9471&  0. 4070 \\
		Proposed    & \textbf{0.9166} 	&\textbf{ 0.7560} 	&\textbf{0.9587} 	&\textbf{0.3610} \\
		
		\bottomrule[.15em]
		
	\end{tabular}

	%}
	
\end{table}

We apply the spatial global average pooling ($ \rm GP_{mean}$) and spatial global standard deviation pooling ($ \rm GP_{std}$) on the extracted feature maps $F_{s}$ to obtain the final quality-aware feature representation $f_{nr}$:
\begin{equation}
\begin{aligned}
f_{mean} &= { \rm GP_{mean}}(F_s), \\
f_{std} &= { \rm GP_{std}}(F_s), \\
f_{nr} &= { \rm cat}(f_{mean}, f_{std}), \\
\end{aligned}
\end{equation}
where $ \rm cat$ is the concatenation operation.
\subsection{Quality Regression Module}
After extracting quality-aware feature representation by the feature extraction module, we need to map these features to the quality scores with a regression model. In this paper, we use two fully connected (FC) layers as the regression model to obtain the frame-level quality due to its simplicity. The two FC layers consist of 128 and 1 neurons respectively. Therefore, we can obtain the frame-level quality score via

\begin{equation}
\begin{array}{c}
q_t = W_{FC}(f), f \in \{f_{fr}, f_{nr}\},
\end{array}
\end{equation}
where $ W_{FC} $ denotes the function of the two FC layers and $q_t$ is the quality of the $t$-th frame, and $t \in [1,2,3,..,N]$.

\subsection{Quality Pooling Module}

\begin{table}
	\small
	%\footnotesize
	\centering
	\renewcommand{\arraystretch}{1.15}
	\caption{The performance of the proposed NR VQA model and the other compared models on the validation set of the Compressed UGC VQA database.}
	\label{NR_Compressed}
	%\resizebox{1\textwidth}{!}{
	\begin{tabular}{c|cccc}
		\toprule[.15em]
		
		Methods & SROCC & KROCC & PLCC & RMSE   \\
		\hline
		NIQE \cite{mittal2012making}     &0.5929  &0.4173  & 0.5977  &0.9764 \\
		
		BRISQUE \cite{mittal2012no}  &0.8346  &0.6477  &0.8886 &0.5587 \\
		
		RAPIQUE \cite{tu2021rapique}  &0.8647  &0.6898  &0.9247&0.4742 \\
		
		VSFA \cite{li2019quality} &0.8925  &0.7280  &0.9587 &0.3463   \\
		
		Proposed    & \textbf{0.9352} 	&\textbf{ 0.7937} 	&\textbf{0.9826} 	&\textbf{0.2260} \\
		
		\bottomrule[.15em]
		
	\end{tabular}	
	%}
	
\end{table}

%Since human is sensitive to drops in video quality and is dull to improvements in video quality, which is called temporal hysteresis effect
Since human is sensitive to drops in video quality and is dull to improvements in video quality, the contribution of the quality of each frame to the overall quality is different. So, we use the quality pooling module to assign the weights to different frames to represent their contribution to the overall quality. In this paper, we use the subjectively-inspired temporal pooling strategy proposed in \cite{li2019quality}, which considers the memory effect of previous frames and the hysteresis effect of next frames to the current frame. The detailed introduction of the subjectively-inspired temporal pooling strategy can refer to \cite{li2019quality}. Here, we denote it as $\rm TP$.
%First, to describe the intolerance to poor quality frames, a memory quality score for previous quality scores It at the t-th frame is defined as the minimum of quality scores over the previous frames.
%\begin{equation}
%\begin{array}{ll}
%l_{t}=q_{t}, & \text { for } t=1, \\
%l_{t}=\min _{k \in V_{\text {prev }}} q_{k}, & \text { for } t>1,
%\end{array}
%\end{equation}
%
%Then, we assign the large weights to worse quality frames among next several frames to describe the effect of hysteresis effect, which is defined as:
%\begin{equation}
%\begin{aligned}
%m_{t} &=\sum_{k \in V_{\text {next }}} q_{k} w_{t}^{k}, \\
%w_{t}^{k} &=\frac{e^{-q_{k}}}{\sum_{j \in V_{\text {next }}} e^{-q_{j}}}, k \in V_{\text {next }},
%\end{aligned}
%\end{equation}
Then, the overall video quality is calculated as
\begin{equation}
\begin{aligned}
q_{t}^{\prime}  &= {\rm TP}(q_t), \\
Q  &= \frac{1}{N} \sum_{t=1}^{N} q_{t}^{\prime}, \\
\end{aligned}
\end{equation}
where $q_{t}^{\prime}$ is the quality of the $t$-th frame which considers the memory effect and the hysteresis effect. $Q$ is the video quality score predicted by the proposed model.

The Euclidean distance is used as the loss function of the proposed model:
\begin{equation}
L=\left\|Q-Q_{label} \right\|^{2},
\end{equation}
where $Q_{label}$ is the ground-truth quality score obtained from subjective experiments.

\section{Experimental Validation}

\subsection{Experimental Protocol}
\subsubsection{Database}
The proposed model is mainly validated on the Compressed UGC VQA database, which is provided by the ICME 2021 Grand Challenge \cite{ugc2021challenge}. The database consists of 6,400 video clips for training, 800 video clips for validation, and 800 video clips for the test. The reference video is compressed into 7 distorted ones by H.264/AVC with different CRF settings. Each video clip is rated by at least 50 subjects.

Besides the Compressed UGC VQA database, we also validate the proposed NR VQA model on two in-the-wild UGC VQA databases, KoNViD-1k\cite{hosu2017konstanz} and LIVE-VQC\cite{sinno2018large}. The KoNViD-1k database includes 1,200 public-domain videos sampled from the YFCC100M database, and is annotated by 642 subjects on the crowdsourcing platform. The LIVE-VQC database consists of 585 in-the-wild videos, which are labeled by 4,776 subjects on the crowdsourcing platform.

\subsubsection{Evaluation Criteria}
\vspace{-0.2cm}
Four common criteria are adopted to evaluate the performance of VQA models, which are Spearman Rank Order Correlation Coefficient (SROCC), Kendall Rank Order Correlation Coefficient (KROCC), Pearson Linear Correlation Coefficient (PLCC) and Root Mean Squared Error (RMSE). SROCC and KROCC indicate the prediction monotonicity of the VQA algorithm, PLCC reflects the prediction linearity, and RMSE represents the predication accuracy. Before calculating the evaluation criteria, we follow the same procedure in \cite{li2019quality} to map the objective score to the subject score using a four-parameter logistic function.

\subsubsection{Experimental Settings}

For the Compressed UGC VQA database, we extract one frame every half second for the feature extraction, and for the KoNViD-1k and LIVE-VQC databases, we extract one frame every second. We use ResNet50 \cite{he2016deep} as the backbone for both the FR and NR feature extraction networks. The weights of the backbone are initialized by training on ImageNet, and other weights are randomly initialized. For all databases, we first resize the resolution of the minimum dimension of videos as 520 while maintaining their aspect ratios. In the training stage, the input video is randomly cropped with a resolution of 448$\times$448, and in the test stage, each video is cropped at the same resolution of 448$\times$448 at the center. The Adam optimizer with the initial learning rate 0.00001 and batch size 6 is used for training the proposed model on a server with NVIDIA V100.

Since there are no training and validation splits for the KoNViD-1k and LIVE-VQC databases, we randomly split the databases into the training set with 80\% videos and the validation set with 20\% videos for 10 times, and report the average values of four evaluation criteria as the final result.

\begin{table}
	\small
	%\footnotesize
	\centering
	\renewcommand{\arraystretch}{1.15}
	\caption{The performance of the proposed NR VQA model and the compared models on the KoNViD-1k database.}
	\label{NR_KoNViD-1k}
	%\resizebox{1\textwidth}{!}{
	\begin{tabular}{c|cccc}
		\toprule[.15em]
		
		Methods & SROCC & KROCC & PLCC & RMSE   \\
		\hline
		
		BRISQUE \cite{mittal2012no}    &0.6567  &0.4761  & 0.6576  &0.4813  \\

		V-BLIINDS \cite{saad2014blind}  &0.7101  &0.5188  &0.7037&0.4595 \\
		
		TLVQM \cite{korhonen2019two} &0.7729  &0.5770  &0.7688 &0.4102   \\
		
		VIDEVAL \cite{tu2021ugc} &0.7832  &0.5845  &0.7803 &0.4026   \\

		Proposed    & \textbf{0.8134} 	&\textbf{ 0.6201} 	&\textbf{0.8143} 	&\textbf{0.3695} \\
		
		\bottomrule[.15em]
		
	\end{tabular}	
	%}
	\vspace{-0.2cm}	
\end{table}

\begin{table}
	\small
	%\footnotesize
	\centering
	\renewcommand{\arraystretch}{1.15}
	\caption{The performance of the proposed NR VQA model and the compared models on the LIVE-VQC database.}
	\label{NR_LIVE-VQC}
	%\resizebox{1\textwidth}{!}{
	\begin{tabular}{c|cccc}
		\toprule[.15em]
		
		Methods & SROCC & KROCC & PLCC & RMSE   \\
		\hline
		
		BRISQUE \cite{mittal2012no}    &0.5925  &0.4162  & 0.6380  &13.1004  \\

		V-BLIINDS \cite{saad2014blind}  &0.6939  &0.5078  &0.7178&11.7659 \\
		
		TLVQM \cite{korhonen2019two} & \textbf{0.7988}  &\textbf{0.6080}  &\textbf{0.8025} &\textbf{10.1454}   \\
		
		VIDEVAL \cite{tu2021ugc} &0.7522  &0.5639  &0.7514 &11.1004   \\

		Proposed    &0.7466 	&0.5518	&0.7815	&13.5143 \\
		
		\bottomrule[.15em]
		
	\end{tabular}	
	%}
	\vspace{-0.2cm}	
\end{table}

\subsection{Experimental results}
\textbf{Full Reference.}
We list the performance of the proposed FR VQA model and other state-of-the-art FR I/VQA models on the validation set of the Compressed UGC VQA database in Table \ref{FR_Compressed}. From Table \ref{FR_Compressed}, it is seen that the proposed model achieves the best performance among all compared models, which demonstrates the effectiveness of the proposed model. We notice that some FR IQA models such as LPIPS and DISTS also have a certain ability to evaluate the quality of the compressed videos since they both extract features from the intermediate layers of the CNN network. C3DVQA uses the 3D CNN network to learn the spatial-temporal features of video, whose performance is superior to other FR IQA models but is interior to the proposed model. %What's more, our FR VQA model won first place on the DMOS track of the ICME Grand Challenge among 6 participants \cite{ugc2021challenge}.

\textbf{No Reference.}
The performance of the proposed NR VQA model and the compared methods on the validation set of the Compressed UGC VQA database is listed in Table 2 and the performance on the KoNViD-1k and LIVE-VQC databases are listed in Table 3 and 4 respectively. From Table 2, we observe that the proposed NR VQA model outperforms the compared NR I/VQA models by a large margin, which indicates the proposed NR VQA model has a strong ability to evaluate the quality of the compressed UGC videos.
% Also, our NR VQA model won second place on the MOS track of the ICME Grand Challenge among 12 participants, and achieves the highest performance on evaluating the quality of the reference videos in the test set of the Compressed UGC VQA database, which is the most common usage scenario in practical applications \cite{ugc2021challenge}.
From Table 3 and Table 4, we find that the proposed model performs the best on the KoNViD-1k database, and is just inferior to TLVQM and VIDEVAl on the LIVE-VQC databases, which indicates that the proposed model is also suitable for evaluating the quality of in-the-wild UGC videos. Since our model does not extract the motion features of the videos and the videos in the LIVE-VQC database usually contain many camera motion or object motion scenes, our model does not achieve the best on the LIVE-VQC database.

\section{Conclusion}
\vspace{-0.2cm}
In this paper, we propose an effective deep learning based VQA framework, which is suitable for both FR and NR VQA models. The proposed VQA framework contains three modules, which are the feature extraction module, quality regression module, and quality pooling regression module. For the feature extraction module, we incorporate the features from the intermediate layers of the CNN model into the final quality-aware feature representation, of which the structure and texture similarity of feature maps are calculated as the feature representation for the FR VQA model, and the staircase structure is adopted to hierarchically add the feature maps from intermediate layers into the final feature maps and their global mean and standard deviation are calculated as the feature representation for the NR VQA model. For the quality regression module, we use the FC layer to regress the quality-aware features into the frame-level score. Finally, a subjectively-inspired temporal pooling strategy is adopted to obtain the final video score. The proposed model achieves the best performance on the Compressed UGC VQA database when compared with other state-of-the-art VQA models and also achieves pretty good performance on the in-the-wild UGC VQA databases.

\vspace{-0.3cm}

\small
\bibliographystyle{IEEEbib}
\bibliography{icme2021template}

\end{document}